\documentstyle[aps,psfig]{revtex}
\def\be{\begin{equation}}
\def\ee{\end{equation}}
\def\bea{\begin{eqnarray}}
\def\eea{\end{eqnarray}}

\begin{document}
\title
{Critical exponents of the random-field ${\rm O}(N)$ model}
\author
{D.E. Feldman}
\address
{Condensed Matter Physics Department,
Weizmann Institute of Science, 76100 Rehovot,
Israel\\ and
Landau Institute for Theoretical Physics,
142432 Chernogolovka, Moscow region,
Russia}
\maketitle
\begin{abstract}
The critical behavior of the random-field Ising model
has been a puzzle for a long time. Different theoretical
methods predict that the critical exponents of the
random-field ferromagnet in $D$ dimensions are the same as in the
pure $(D-2)$-dimensional ferromagnet with the same number of the
magnetization components.
 This result
contradicts the experiments and simulations.
We calculate the critical exponents of the random-field
O$(N)$ model with the $(4+\epsilon)$-expansion
and obtain values different from the critical exponents
of the pure ferromagnet in $2+\epsilon$ dimensions.
In contrast to the previous approaches we take into account an
infinite set of relevant operators emerging in the problem.
We demonstrate how these previously missed relevant operators
lead to the breakdown of the $(6-\epsilon)$-expansion for the
random-field Ising model.
\end{abstract}
\pacs{75.10.Nr, 64.60.Fr, 05.70.Jk}

The role of quenched disorder in condensed matter depends on
its strength but even weak disorder can strongly modify
the nature of the phase transition. This effect is most prominent
in the case of random-field (RF) disorder which breaks both the translational
symmetry and the symmetry with respect to transformations of the order
parameter. In particular, in some systems the arbitrarily weak disorder
of this type destroys long-range order \cite{review}. The strong
effect of weak disorder makes it difficult to apply the standard
perturbative methods of phase transition theory to RF
problems.  Besides, it is much more difficult to solve exactly a random
model than a pure one. As a result, a theory of phase transitions in
the presence of random fields is still absent.

The large number of systems with RF disorder provides a strong
motivation to develop such theory. Some of these systems are known for a
long time. Examples are disordered antiferromagnets in the external
magnetic field \cite{antiferro}, binary liquids in random porous media
\cite{binliq} and vortices in disordered superconductors \cite{XYreview}.
Recently a lot of attention was devoted to disordered liquid crystals
\cite{liqcry} and liquid He-3 in aerogels \cite{He3}.
In contrast to the pure systems even the question
of ordering at low temperatures in the presence of random fields
is nontrivial. The more difficult problem of the critical
behavior is still open in spite of two decades of investigations.

After a lot of controversy the lower critical dimension
of the RF Ising model was found \cite{review} but the structure
of the phase diagram and the nature of the transition to
the ferromagnetic state are still unclear.
Different theoretical approaches to the paramagnet-ferromagnet
transition predict the dimensional reduction \cite{review}:  the critical
exponents of the RF O$(N)$ model in $D$ dimensions should be
the same as in the pure O$(N)$ model in $D-2$ dimensions.  This
prediction relates the three-dimensional RF Ising model to the
one-dimensional pure system in contradiction to the presence of
long-range order in the former and its absence in the later.  Moreover,
the high-temperature expansion \cite{hT}
shows that the dimensional reduction rule is invalid in any dimension
less than the upper critical dimension 6.  The most elegant derivation
of the dimensional reduction \cite{PS} provides an (incorrect) exact
solution of the RF O$(N)$ model at zero temperature. The
failure of that approach was explained by Parisi on the basis of the
complicated energy landscape of the model \cite{LeH}. However, such
explanation is insufficient for the renormalization group (RG) and
$1/N$-expansion. It was conjectured \cite{LeH} that RG fails because
of non-perturbative corrections to the expansion in $\epsilon$. However,
such corrections are not found and recent numerical results \cite{hT}
suggest that the existing predictions for the coefficients
of the series in $\epsilon$ are wrong \cite{footnote}.

A possible reason why the perturbation theory does not
provide satisfactory results is the appearance of some additional
relevant operators missed by the existing approaches.
This phenomenon is responsible
for the failure of the standard RG theory to predict the
order of the phase transition in some systems \cite{Potts}.
The possibility of a similar phenomenon in the RF
O$(N)$ model was first suggested in Ref. \cite{Fisher}
but no new values of the critical exponents were found. 
Recently a similar idea was used in Ref. \cite{BD}
but it also did not allow to calculate the exponents.
Besides, the approach \cite{BD}
involves delicate manipulations with the replica limit 
and its validity is not clear \cite{critique}.
In the present paper we demonstrate that there are indeed some additional
relevant operators in the problem although not those found in Ref. \cite{BD}.
We calculate the critical exponents of the RF
O$(N)$ model in $4+\epsilon$ dimensions and show that
they do not satisfy the dimensional reduction.
We also demonstrate how the additional relevant operators lead to the
breakdown of RG in $6-\epsilon$ dimensions.
Some phenomenological approaches \cite{phen} allowed one to obtain
the critical exponents different from the dimensional reduction
prediction. However, their results based on different
unproven assumptions contradict each other.
An important
breakthrough was made by Mezard and Young \cite{MY}
who considered the possibility of the replica symmetry breaking
in the RF O$(N)$ model at large $N$.
Unfortunately,  the approach \cite{MY}
did not allow the calculation of the
exponents. The present paper contains the first systematic method that
explains the failure of RG and allows us to find the critical
exponents of the RF systems.

In the critical point the RF O$(N)$ model can be described
by the connected and disconnected correlation functions \cite{review}:
\be
\label{0}
G_{\rm con}({\bf q})=
\langle[{\bf n}({\bf q}){\bf n}(-{\bf q})]
- [{\bf n}({\bf q})]
[{\bf n}(-{\bf q})]\rangle\sim q^{-2+\eta};
G_{\rm dis}({\bf q})=
\langle[{\bf n}({\bf q}){\bf n}(-{\bf q})]\rangle\sim
q^{-4+\bar\eta},
\ee
where ${\bf n}({\bf q})$ is the Fourier
component of the magnetization, the
square brackets denote the thermal average and the
angular brackets denote the disorder average.  To calculate the critical
exponents $\eta$ and $\bar\eta$ we develop the $\epsilon$-expansion
near the lower critical dimension 4
following the line of Ref.  \cite{Feldman}. Our starting point is the
Hamiltonian of the RF O$(N)$ model
\be
\label{1}
H=\int d^D x[J\sum_\mu\partial_\mu{\bf n}({\bf x})
\partial_\mu{\bf n}({\bf x})
- \sum_k ({\bf h}_k({\bf x}){\bf n}({\bf x}))^k],
\ee
where the unit vector of the magnetization
${\bf n}({\bf x})$ has $N$ components and ${\bf h}_k$ are
random fields with zero average, $\langle h_{k,\alpha}({\bf x})
h_{q,\beta}({\bf y})\rangle=H_k\delta_{\alpha\beta}\delta_{kq}
\delta({\bf x}-{\bf y})$.
The Hamiltonian includes an infinite set of random
anisotropies of different ranks. These contributions are allowed by
symmetry and turn out to be relevant in the RG sense. The replica
Hamiltonian has the form
\be
\label{2}
H_R=\int d^D x[\sum_a\frac{1}{2T}\sum_\mu
\partial_\mu{\bf n}_a\partial_\mu{\bf n}_a - \sum_{ab}\frac{R({\bf
n}_a{\bf n}_b)}{T^2}+\dots],
\ee
where $a,b$ are replica indices, $R(z)$ is some function, $T$ the
temperature and the dots denote the irrelevant terms. 
Near the zero-temperature fixed point the whole function $R(z)$ is
relevant. 
Indeed,  to ensure
the fixed length condition
${\bf n}_a^2=1$ at each RG step, 
we ascribe the scaling dimension 0
to the magnetization ${\bf n}$.
The scaling dimension of the temperature
is $-2+O(\epsilon)$. 
The relevance of any operator
is determined by the number of the derivatives in it
and the power in which it contains the temperature.
This shows that all operators 
$R_k=\sum_{ab}({\bf n}_a{\bf n}_b)^k/T^2$
are relevant in the same space dimensions.
The functional RG analysis
in $4+\epsilon$ dimensions repeats the same steps as
the $(4-\epsilon)$-expansion developed in Ref. \cite{Feldman} and
results in the following RG equation for the disorder strength
\begin{eqnarray}
0=\frac{dR(\phi)}{d \ln L}=-\epsilon R(\phi) + (R''(\phi))^2 - 2R''(\phi)
R''(0) -  & & \nonumber\\
\label{3}
(N-2)[4R(\phi)R''(0)+2{\rm ctg}\phi R'(\phi) R''(0) -
\left(\frac{R'(\phi)}{\sin\phi}\right)^2] + O(R^3,T), & &
\end{eqnarray}
where we define $\cos\phi={\bf n}_a{\bf n}_b$ to
make the equation more compact.
The RG equation is valid in the zero-temperature fixed point
that describes the phase transition in the RF ferromagnet.
The critical exponents (\ref{0}) can be expressed \cite{Feldman} via the
RG charge $R(\phi)$:
\be
\label{4}
\eta=-2R''(0);\bar\eta=-2(N-1)R''(0)-\epsilon.
\ee

Note that $R''(0)<0$ since $G_{\rm con}({\bf r})\sim r^{2(N-1)R''(0)}$
must be limited.
The solution of Eq. (\ref{3}) can be found numerically
with shooting. Equation (\ref{3}) has only a discrete set of
the solutions which are everywhere finite.
The region of possible initial conditions
$R''(0)$ is limited by the restrictions
\be
\label{5}
\epsilon/2(N-3)\ge -R''(0)=\eta/2 \ge \epsilon/2(N-2).
\ee
The first inequality follows from the Schwartz-Soffer inequality
\cite{SwSo} and is the stability condition for the fixed point
 \cite{Feldman}.
The second one can be derived from the third term of the
expansion of the function $R(\phi)$ at small $\phi$:
\be
\label{expnas}
R(\phi)=-\frac{(N-1)R''(0)^2}{\epsilon+4(N-2)R''(0)}
+\frac{R''(0)}{2}\phi^2
\pm\sqrt{\frac{R''(0)\epsilon+2(N-2)R''(0)^2}{18(N+2)}}|\phi|^3+\dots
\ee
Since coefficients of Eq. (\ref{3}) are singular at
$\phi=0,\pi$, one has to use the expansions of $R(\phi)$ in powers of
$|\phi|, (\pi-\phi)$ near those points at solving numerically
the RG equation.
Note that the function $R(\phi)$
is non-analytical at $\phi=0$.
We have solved the RG equation at $N\le 5$.
It turns out that for
$5\ge N \ge 3$ Eq. (\ref{3}) has only one solution compatible with the
condition (\ref{5}). The critical exponents calculated for those $N$ are
given in Table I.
One can see that the critical exponents are different from the
dimensional reduction values $\eta(4+\epsilon)=\bar\eta(4+\epsilon)
=\eta_{\rm pure}(2+\epsilon)=\epsilon/(N-2)$, where
$\eta_{\rm pure}$ is taken from Ref. \cite{ZJ}.

The second of the inequalities (\ref{5}) 
shows that
for the XY model $(N=2)$ there is no appropriate solution of
the RG equation. This can be explained by the fact
\cite{XYreview} that the
RF XY model possesses quasi-long-range order
in dimension 3. Hence, it has a phase transition in 3 dimensions
and thus its lower critical
dimension near which the $\epsilon$-expansion should be developed
is less than 4.

It was claimed \cite{Young} that
in the first order in $\epsilon$ the critical
exponents of the RF ferromagnet in $4+\epsilon$
dimensions are the same as in the pure O$(N)$ model in
$2+\epsilon$ dimensions. That prediction could be obtained
from Eq. (\ref{3}) with the wrong assumption
that there is an analytical solution $R(\phi)$
controlling the critical point.

We have demonstrated the existence of an infinite set of relevant operators
near 4 dimensions in the RF O$(N)$ model.
A possible explanation of the failure of the 
($6-\epsilon$)-expansion 
is thus the appearance of additional relevant
operators below some dimension $D<6$. However,
we show that another scenario takes place:
an infinite set of relevant operators
emerges in any dimension less than
the upper critical dimension 6.
Since the upper critical dimension is the
same for the Ising and O$(N)$ models our approach allows us to
consider both cases in the same way.

Near 6 dimensions it is convenient to use the Ginzburg-Landau
model with the random field ${\bf h}({\bf x})$
\be
\label{6}
H=\int d^Dx [(\nabla{\bf m})^2+g({\bf m}^2)^2-{\bf h}(\bf x){\bf m}]
\ee
that can be described by the replica Hamiltonian
\be
\label{7}
H_R=\int d^Dx [\sum_a(\nabla{\bf m}_a)^2+\sum_a g({\bf m}_a^2)^2-
\Delta\sum_{ab}{\bf m}_a{\bf m}_b],
\ee
where ${\bf m}_a$ are the replicas of the $N$-component magnetization.
The standard power counting suggests that
all operators which are relevant in $6-\epsilon$ dimensions
are included into Eq. (\ref{7}).
However, if it were so the dimensional reduction would be
correct. The hint why the power counting
fails is given by the theory of the metal-dielectric
transition. It was argued that in the nonlinear
sigma-model for the metal-dielectric transition
one should include an infinite set of relevant operators
\cite{KLY} missed by the power counting.
For any eigen-operator $A_k$ of the RG transformation
the RG equation has the structure
\be
\label{8}
\frac{d A_k}{d\ln L}=(C_0^k+C_1^k\epsilon^{s}+o(\epsilon^{s}))
A_k + O(A^2).
\ee
The power counting is based on the sign of the constant $C_0$
and predicts that the operator $A_k$ is irrelevant as $C_0^k<0$.
However, if the ratio $C_1^k/C_0^k$
grows up to infinity at large $k$
then for any fixed $\epsilon$ there is such $k$ that the correction
$C_1^k\epsilon^s$ is greater than $C_0^k$. Hence, the role of the operators
$A_k$ depends on the signs of $C_1^k$.
In the $\phi^4$-theory without the random field this sign is negative for
any operator which is expected to be irrelevant from
the power counting \cite{Wegner}. This agrees with the success of
the RG approach in that problem. On the other hand,
we shall see that the signs
of these constants are positive for some operators in the
RF problem. This signals that additional relevant operators
emerge.

For the pure $\phi^4$-model
one can define the rank of an operator with $q$ derivatives and
the $p$th power of the order parameter
as $r=p+q-4$. Since only the operators of the same rank
can mix \cite{ZJ}, for the calculation of the anomalous dimension
$C_1^k\epsilon^s$ of an operator with rank $r_k$
one can ignore the diagrams which produce the
operators of the higher ranks.
It is easy to check that the definition of the rank
has to be modified in the RF problem to ensure
that the operators of different ranks do not mix:
$r=p+q+2t-6$, where $t$ is the number of the different replica indices
in the operator.

To avoid the difficult problem of operator mixing we
guess a set of relevant operators which do not mix with the other
operators of the same rank
up to the second order in $\epsilon$.
It turns out that the following operators are relevant and do not
mix with the other:
$A_k=\sum_{ab}[({\bf m}_a-{\bf m}_b)^2]^k$.
At $k>1$ their canonical dimensions
\be
\label{cd}
d_c^k=(4-\epsilon)k - 6+\epsilon
\ee
are positive and proportional to
$k$. We shall see that the anomalous dimensions are negative and
proportional
to $(k\epsilon)^2$. If one imposes the fixed length restriction
${\bf m}^2=1$ so that the Ginzburg-Landau model reduces to
the O$(N)$ model then
the operators $A_k$ can be represented as the
linear combinations of the random anisotropies
$B_k=\sum_{ab}({\bf m}_a{\bf m}_b)^k$. This is natural since
the operators $B_k$ are relevant
in $4+\epsilon$ dimensions as one could see above.

To the first order in the quartic vertex $g$
the anomalous dimensions of the operators $A_k$ are zero.
To prove this we demonstrate that
all diagrams with one vertex $gm_a^4$ and one vertex $A_k^{ab}=[({\bf m}_a-{\bf m}_b)^2]^{k}$
either are equal to zero or produce
operators of higher ranks. 
Indeed, any line of the Feynman diagram corresponds to
the product of a momentum-dependent factor and two operators
which differentiate the vertices at the ends of the line
with respect to replicas ${\bf m}_a$ of the magnetization.
For example, line 2 in Fig. 1 acts on the vertex $A_k^{ab}$
as the differential operator
$\partial_2=\partial/\partial {\bf m}_a$.
Line 1 in which the vertex $\Delta$ is inserted 
acts on the vertex $A_k^{ab}$ as the operator
$\partial_1=\partial/\partial{\bf  m}_a+\partial/\partial{\bf m}_b$.
Obviously, $\partial_1 A_k^{ab}=0$.
Hence, any diagram including line 1 is equal to zero.
On the other hand, any diagram with one vertex $g$, one vertex $A_k$,
and without lines in which $\Delta$ is inserted
produces an operator of a higher rank.
Fig. 2 shows the only nonzero diagram of
the order $g^2$ that does not increase the rank of the operators $A_k$.
Calculating this diagram
one obtains the anomalous dimension
\be
\label{9}
d_{\rm an}^{k,N}=k(N+2)\epsilon^2/2(N+8)^2 - \epsilon^2
[Nk(2k+1)+16k^2-10k]/2(N+8)^2
\ee

The anomalous dimension (\ref{9}) is negative and proportional to $k^2$.
Thus, at any fixed $\epsilon$ one expects that the operators $A_k$
with $1/\epsilon^2 \lesssim k$ 
are relevant but missed by the existing
theoretical methods. Although the rigorous analysis requires
consideration of all orders in $\epsilon$, the appearance
of an infinite set of relevant operators in
$6-\epsilon$ dimensions is plausible
since such set exists near 4 dimensions and since
the alternative explanation \cite{LeH} of the failure of RG
due to non-perturbative corrections is hardly
compatible with the existing numerical results \cite{hT}.

In conclusion, we have calculated critical exponents
of the O$(N)$ model in $4+\epsilon$ dimensions and demonstrated
that they do not obey the dimensional reduction. The failure
of the dimensional reduction is related to the appearance
of an infinite set of relevant operators.

I thank Y. Gefen, I.V. Lerner, V.L. Pokrovsky, and M.G. Stepanov for
useful discussions and R. Whitney for the critical reading of the manuscript.
This work was supported by the Koshland fellowship and RFBR grant
No. 00-02-17763.

\newpage

\begin{table}
\caption{Critical exponents of the RF O$(N)$ model. $\eta$ and 
$\bar\eta$ are the exponents of the connected and disconnected
correlation functions respectively.}
\label{table1}
\begin{tabular}{cccc}
$N$        & 3    & 4    & 5    \\
\tableline
$\eta$     &  $5.5\epsilon$ & $0.79\epsilon$ & $0.42\epsilon$ \\
$\bar\eta$ &  $10.1\epsilon$ & $1.4\epsilon$ & $0.70\epsilon$\\
\end{tabular}
\end{table}

\begin{figure} \label{fig1}
  \hfill
  \psfig{file=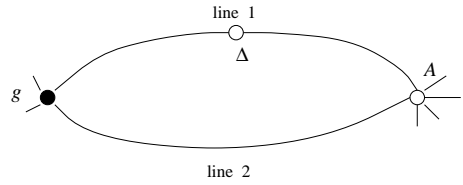,width=2.9in,angle=270}
  \hfill\hfill
\caption{A first order diagram contributing to the anomalous dimension.}
\end{figure}

\begin{figure} \label{fig2}
  \hfill
    \psfig{file=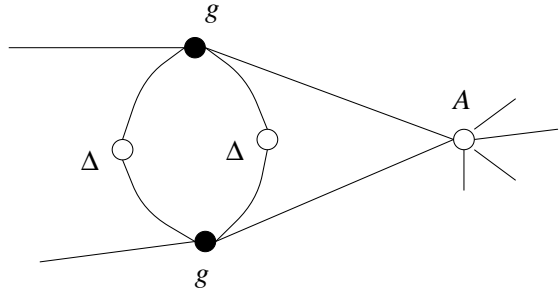,width=2.9in,angle=270} 
  \hfill\hfill
\caption{A second order diagram contributing to the anomalous dimension.}
\end{figure}

\end{document}